\documentclass{IEEEtran}

\usepackage{graphicx}
\usepackage{subcaption}
\usepackage{amsmath}
\usepackage{nccmath}
\usepackage{cite}
\usepackage{xcolor}
\usepackage{adjustbox,lipsum}
\pdfoutput=1

\begin{document}

\title{Deep UL2DL: Data-Driven Channel Knowledge Transfer from Uplink to Downlink}

%
%


\author{Mohammad Sadegh~Safari, Vahid~Pourahmadi, and Shabnam~Sodagari,~\IEEEmembership{Senior Member, IEEE}
\thanks{Mohammad Sadegh~Safari and Vahid~Pourahmadi (corresponding author) are with the Electrical Engineering Department, Amirkabir University of Technology, Tehran, Iran, (email: v.pourahmadi@aut.ac.ir)}
\thanks{Shabnam Sodagari is with the Electrical Engineering Department, California State University, Long Beach, CA 90840 USA, (e-mail: shabnam@csulb.edu).}
}

\maketitle



\begin{abstract}
	Knowledge of the channel state information (CSI) at the transmitter side is one of the primary sources of information that can be used for efficient allocation of wireless resources. Obtaining downlink (DL) CSI in Frequency Division Duplexing (FDD) systems from uplink (UL) CSI is not as straightforward as in TDD systems. Therefore, users usually feed the DL-CSI back to the transmitter. To remove the need for feedback (and thus having less signaling overhead), we propose to use two recent deep neural network structures, i.e., convolutional neural networks and generative adversarial networks (GANs) to infer the DL-CSI by observing the UL-CSI. The core idea of our data-driven scheme is exploiting the fact that both DL and UL channels share the same propagation environment. As such, we extracted the environment information from UL channel response to a latent domain and then transferred the derived environment information from the latent domain to predict the DL channel. 
To overcome incorrect latent domain and the problem of oversimplistic assumptions, in this work, we did not use any specific parametric model and instead used data-driven approaches to discover the underlying structure of data without any prior model assumptions. To overcome the challenge of capturing the UL-DL joint distribution, we used a mean square error-based variant of the GAN structure with improved convergence properties called boundary equilibrium GAN (BEGAN). For training and testing we used simulated data of Extended Vehicular-A (EVA) and Extended Typical Urban (ETU) models. Simulation results verified that our methods can accurately infer and predict the downlink CSI from the uplink CSI for different multipath environments in FDD communications. 
\end{abstract}

\section{Introduction} 
\label{ssec:intro}

One key feature of newer generations of cellular networks is their efficient use of frequency bands and energy. To achieve this goal, they use various techniques, such as water-filling, appropriate precoding and beamforming. In most of these techniques, the Channel State Information (CSI) should be available at the transmitter side (CSIT). In Time Division Duplexing (TDD) systems, Up-Link (UL) and Down-Link (DL) frequencies are equal, so we can use channel reciprocity and simply infer the DL channel by observing the UL channel. In Frequency Division Duplexing (FDD) systems, however, DL channel and UL channel have different frequencies. Therefore, we cannot use channel reciprocity to infer the DL channel. The most commonly used solution is that the user (receiver) first measures (estimates) the DL channel, and then sends its information back to the transmitter. This solution has two major disadvantages: delay and overhead. If the delay is larger than the coherence time, the actual DL channel is different from what has been fed back by the user. In addition, in new generations of mobile networks the transmitter has a large number of antennas. For example, for a fourth-generation transmitter with 64 antennas, the need to learn the DL channel (pilot transmission and feedback data) consumes a large portion of the transmitter's traffic \cite{ji2017overview}. This very large overhead is a major challenge in LTE networks. \cite{ji2017overview}. 
These challenges have such important effects on the network performance that despite some significant advantages of FDD systems, such as continuous transmission \cite{chan2006evolution}, in recent years, TDD has attracted more attention.

\begin{figure}
	\centering
	\includegraphics[width=3.0in]{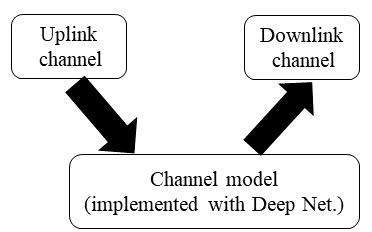}
	\centering
	\caption{UL to DL knowledge transfer procedure: extracting environment information from UL and transferring it to DL}
	\label{fig:model}
\end{figure}

To eliminate the need for the feedback (and so its associated overhead and delay), there are several studies that aim to infer the DL channel by observing the UL channel in FDD systems. DL-CSI estimation methods in
\cite{almosa2017downlink, palleit2011prediction, pedersen1999joint}  are based on  the assumption that the difference between the dominant angle of arrival (AOA) in UL and the dominant angle of departure (AOD) in DL is small and directional properties of UL and DL are correlated. For example, a great deal of measurements have shown that with probability of about 81 percent, this difference is smaller than 4.5 degrees \cite{hugl2002spatial}. Therefore, by having the dominant AOA in UL, the dominant AOD in DL can be obtained and used for purposes like beamforming.

Some works \cite{esswie2017novel,xie2018channel,decurninge2015channel,aste1998downlink} are based on covariance matrix due to channel matrix slow variations. In \cite{esswie2017novel} a transformation matrix is used to convert UL covariance matrix to DL covariance matrix.  \cite{decurninge2015channel} is based on the concept of dictionary learning and it has two phases: training and exploitation. In the training phase, they make a dictionary with corresponding DL and UL covariance matrices (by changing user location, they have constructed the dataset of different input and output pairs). {{In the exploitation phase, by observing the UL covariance matrix, DL covariance matrix is constructed by interpolation of stored dictionary with various methods.}} 

In \cite{vasisht2016eliminating,hu2017channel,han2018efficient} by taking into account the multipath structure of the channel, they extract the paths of the signal independent of frequency and hence can infer channel response in any desired frequency band. For example, in \cite{vasisht2016eliminating}, the authors consider four parameters for every path (path attenuation, path length, an independent phase shift for modeling reflection, and angle of arrival of the path). Then, they try to estimate these parameters using the UL-CSI. The resulting model is then used for the prediction of the DL-CSI.

AOA-based methods are often not usable in cases where the accurate channel response is required and are often used only for beamforming. In path extraction-based methods, we can obtain the accurate channel response at any desired frequency, but such methods are often based on assumptions that may not be practically feasible. For example, path attenuation is considered independent of frequency \cite{vasisht2016eliminating}. This assumption is only true if the difference between DL and UL frequencies is small. In \cite{han2018efficient}, to investigate the dependence of path attenuation on frequency, limited feedback was used and it was verified that deriving the DL channel, with the assumption of frequency-independent path attenuation, is not very accurate. Covariance-based algorithms also depend on different environmental factors, such as the correlation among antennas. However, when antenna correlation is poor, \cite{han2010potential} showed that it is not appropriate to use correlation based methods.

To overcome the disadvantages of the above-mentioned methods, artificial intelligence can be used for channel estimation. In recent years, artificial intelligence has revolutionized human life, so that it was called the fourth industrial revolution. One of the leading areas of artificial intelligence and machine learning is deep learning, which has been very successful in many cases, such as machine vision, speech recognition, and object {detection}. In some cases deep learning even exceeds human performance \cite{lecun2015deep}. Deep learning has been also used in physical layer communications \cite{o2017introduction,o2018approximating,o2018physical,qin2018deep,soltani2018deep,wen2018deep, wang2019ul, yang2019deep}. O'Shea {\it et al.} \cite{o2017introduction} considered a communication system at the physical layer as an autoencoder and designed an end-to-end system that optimizes transmitter and receiver simultaneously in one process. Nevertheless, their method is end-to-end, whereas one advantage of our method is designing a separate channel estimator block. In other wordes, our method can be easily inserted within current systems, without having to replace the whole architecture by an end-to-end design.
In addition, \cite{o2018approximating} used a variational GAN to capture the stochastic model of the channel to learn its probability density function (PDF). In \cite{o2018physical} authors used an adversarial network to model the channel input-output conditional probability. In \cite{soltani2018deep}, a super resolution network cascaded with a denoising autoencoder was used to estimate channel response based on some known pilots. CsiNet introduced in \cite{wen2018deep}, to perform limited CSI feedback in FDD systems, encodes channel response at one side (user) and decodes received feedback with a decoder at the other side (base station).
\textcolor{black}{There are also few works \cite{wang2019ul, yang2019deep} that used deep learning to predict DL-CSI in FDD systems.}

Motivated by such applications, in this paper, we propose a novel method based on deep neural networks that predicts the DL-CSI based on the past UL-CSI measurements. Use of the deep networks {enables} us to expand the search space of the environment propagation model (beyond the current mathematical models) and therefore, it can capture more insights on how to infer DL-CSI from the knowledge of UL-CSI. 

In essence, the core idea of our scheme is that the way that the channel {affects} the transmitted signal (regardless of whether it is UL or DL) is related to the structure of the environment in which the signal is propagating (e.g., the objects, which are in the environment, the materials that they are made of, their shape, etc.). By knowing the fact that both DL and UL channels share the same propagating environment (assuming of course no sudden changes in the environment), we use the data-driven approach to extract the environment information from UL channel response to a latent domain and then transfer the derived environment information from the latent domain to the DL channel, as in Fig. \ref{fig:model}. 

To achieve this goal we use two types of Deep Networks: Convolutional Neural Networks (CNNs) and a specific type of Generative Adversarial Networks (GANs) called  Boundary equilibrium GAN (BEGAN), which is based on the Mean Square Error (MSE). For training and testing  we use simulated data of Extended Vehicular-A (EVA) and Extended Typical Urban (ETU) models. Our results verify the effectiveness of our schemes. 

The main contributions of this paper can be summarized as the follows:
\begin{itemize}
	\item using a latent space for transferring channel information from UL to DL;
	\item exploiting a deep CNN structure to model the UL to DL mapping function; and
	\item solving the problem of DL channel estimation from the UL information by casting it to image inpainting techniques of deep generative networks.
\end{itemize}

Moreover, it is worth mentioning that to fully characterize a DL channel of a  multiple-input multiple-output (MIMO) system, we should characterize a 4-dimensional space, i.e., we should find out the channel effects (on both the amplitude and the phase of the signal) between 1) each transmit antenna and 2) each receive antenna, for 3) each of the subcarriers in our frequency range and for 4) each time slot. In most of the previous studies the prediction of DL matrix is investigated in terms of the MIMO channel matrix  and their aim was not to determine the channel effect in the time-frequency domain  \textcolor{black}{(or if they considered the time-frequency domain they did not investigate the MIMO case \cite{wang2019ul}}.
In this work, instead of looking high level at the transmitter and receiver antennas and giving one value to each pair, we focus on one transmit-receive antenna pair and predict the DL channel over a block of time and frequency. \textcolor{black}{Then, we extend the results to the MIMO case.}

The rest of this paper is organized as follows. In section \ref{sec:background}, we will describe CNNs and GANs as the tools that we used to predict DL channel. Section \ref{sec:problemdefinition} provides detailed discussion on the prediction problem. Section \ref{sec:proposedscheme} contains two approaches for solving the DL prediction problem, i.e., the direct approach and the generative approach. In section \ref{sec:implementation}, we explain implementation details of networks and provide simulation results \textcolor{black}{along with further discussions and insights about the results.} Section \ref{sec:conclusion} draws a conclusion on this paper.

\section{background} \label{sec:background}
In the following subsections we briefly discuss two special types of deep neural networks used in this paper to predict the DL channel.

\subsection{Convolutional Neural Networks}
One of the interesting neural network structures widely used in artificial intelligence (AI) community is the Convolutional Neural Network (CNN). CNNs could have many hidden layers and usually they are one of the three types of convolutional layers, pooling layers, and fully connected layers.
CNN is a powerful tool, specially in analyzing two-dimensional (2D) data like images. It is mainly due to the structure of the convolutional layer, which computes the output by convolving filter (kernel) weights with the input image (data). The value of each point in the output image is equal to the cross-correlation of the filter and the corresponding area in the input image \cite{dumoulin2016guide}. After applying convolution operation on the input data, an activation function will be applied~\textcolor{black}{to increase non-linearity of the network.} The output of the activation function will pass through the next layer as the input. \textcolor{black}{CNNs can deduce the main features of input data, thus removing the need for manual feature extraction. We explain more details about the CNN specifics used in this paper in Section \ref{sec:implementation}.}

\subsection{Generative Adversarial Networks}\label{ssec:GAN}

Generative adversarial networks (GANs) are among the most powerful generative models that capture data distribution \cite{goodfellow2014generative}. They are based on game theory and consist of two networks: the generator and the discriminator, which are trained simultaneously. Considering a noise vector $z$ as the input, (typically with normal distribution) the generator tries to create images similar to real ones while the discriminator tries to distinguish generated images from real ones. Training of GANs is a two-player mini-max game. The generator tries to maximize error probability of discriminator (this means the discriminator assigns high probability of being real to generated images) while the discriminator tries to minimize the probability of being real for generated images. GANs are hard to train and non-convergence or instability are their main problems \cite{salimans2016improved}.

Many different structures have been recently proposed for new implementations of GANs. 
In the original GAN, the output of the discriminator is a positive number between 0 and 1. This number represents the probability that the discriminator input image is in fact a real image (not a generated image). Such a discriminator is the most common type of discriminator networks in GAN's literature. For the first time in energy-based GAN (EBGAN) \cite{zhao2016energy}, an autoencoder was used as the discriminator. \textcolor{black}{An autoencoder is an unsupervised neural network that reduces the dimensionality of the input data by learning the main latent variables of the data, i.e., encoding the data.} In EBGAN, the discriminator objective is to maximize reconstruction error of generated images while minimizing it for real ones. EBGAN
generator's structure is similar to the decoder part of the discriminator. Using an autoencoder as the discriminator makes training easier, faster and more stable. Boundary equilibrium GANs (BEGANs) \cite{berthelot2017began} are improved versions of EBGANs and use the same structure but BEGANs aim to match autoencoder loss distribution instead of matching data distribution directly. To train such networks an equilibrium is
\begin{equation} \label{eqn:beganlib}
\gamma  \times E\left[ {L\left( x \right)} \right] = E\left[ {L\left( {G\left( z \right)} \right)} \right]\,
\end{equation}	
where $L\left( x \right)$ and $L\left( {G\left( z \right)}\right)$ are the autoencoder reconstruction losses when it gets a real image and a generated image, respectively. The vector $z$ is a random vector of size $64$ uniformly sampled between $-1$ and $1$. $E[.]$ represents the expectation operation. In Eq. \eqref{eqn:beganlib}, \(\gamma  \in \left[ {0,1} \right]\) has an inverse relation with the diversity of the generated images, meaning that if  \(\gamma\) is set to a larger number, the generator creates less diverse images.  

In BEGAN, \({L_D}\) and \({L_G}\) denote the discriminator and generator loss or objective functions, respectively,
and they are defined as~\cite{berthelot2017began}

\begin{equation} \label{eqn:eqarray} 
\begin{array}{l}{L_D} = L(x) - {k_t}.L\left( {G\left( {{z}} \right)} \right)\\{L_G} = L\left( {G\left( {{z}} \right)} \right)\\{k_{t + 1}} = {k_t} + {\lambda _k}\left( {\gamma L(x) - L\left( {G\left( {{z}} \right)} \right)} \right)\end{array}
\end{equation}
where \({L_D}\) is the difference between the reconstruction loss of real images and the reconstruction loss of generated images, which is scaled with the parameter \({k_t}\) introduced to maintain Eq. \eqref{eqn:beganlib}. Based on proportional control theory, \cite{berthelot2017began} suggests that \({k_t}\) should be updated using the last equation in \eqref{eqn:eqarray} and  \({\lambda _k}\) is its learning rate.

\textcolor{black}{BEGAN, which is a modification of GAN, converges by using the parameter $\gamma$ for controlling the equilibrium between the generator and discriminator so that neither wins over the other. The discriminator is an autoencoder that updates its weights to reconstruct real images with minimum loss $\mathcal{L}(x)$, while simultaneously increasing the reconstruction loss of images produced by the generator $\mathcal{L}(G(z))$. In contrast, the objective of the generator is to minimize the reconstruction loss of generated images $\mathcal{L}(G(z))$. The value of $\gamma \in [0,1]$ determines the level of emphasis on each of these two losses. Lower values of $\gamma$ mean higher emphasis on minimizing the reconstruction loss of generated images than the cost of real images. This forces the generator to produce more realistic images. This ratio is dynamically adapted using $k_t$, which is updated over time as in Eq. (\ref{eqn:eqarray}). The parameter $\lambda$ has a similar interpretation as the learning rate (for example, in algorithms such as gradient descent) and is usually set to $0.001$.}

Visual inspection is typically  the only way to determine convergence in GANs but a convergence measure can also be defined \cite{berthelot2017began} as
\begin{equation} \label{eqn:conergence}
{M_{Global}} = L(x) + \left| {\gamma L(x) - L(G({z}))} \right|.
\end{equation}

In \eqref{eqn:conergence}, smaller \({M_{Global}}\) is desired  as it means both smaller reconstruction loss for real images and maintaining Eq. \eqref{eqn:beganlib}.

\section{Problem Definition and Formulation} \label{sec:problemdefinition}

\begin{figure}
	\centering
	\includegraphics[scale=0.8]{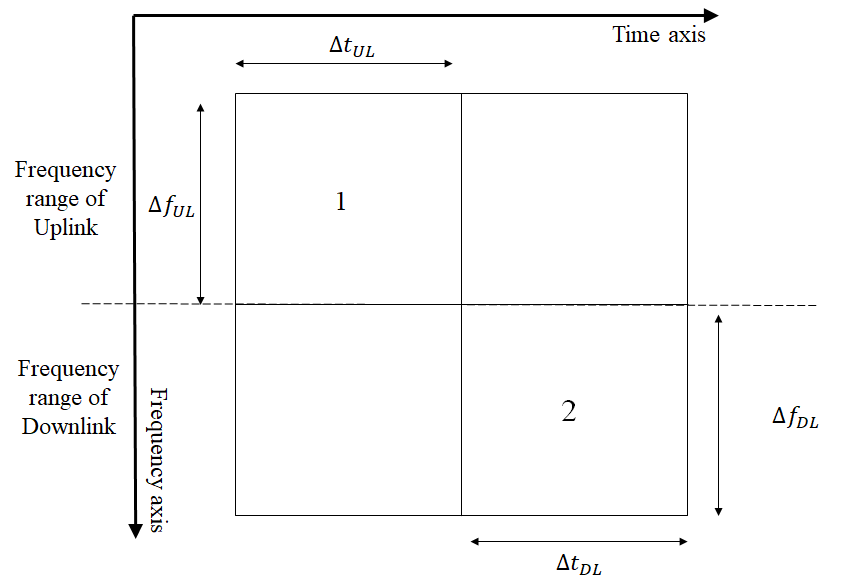}
	\centering
	\caption{UL-DL joint frame structure}
	\label{fig:frame}
\end{figure}

Consider a network composed of a base-station (transmitter) and a user (receiver). To increase the network spectral efficiency (by techniques such as water-filling and beamforming in the case of multiple antennas), the base-station needs to know the DL-CSI. When a user sends its data on the UL channel for example, if it uses OFDM method, it allocates some of its subcarriers and time-slots to pilots transmission. Using the pilots, the base-station can estimate UL-CSI, but in an FDD system UL and DL CSI reciprocity does not hold. So, one way to get that information is to first send pilots in the downlink and after the user estimates the DL-CSI, it sends  DL-CSI over the feedback link. However, this scheme leads to higher overhead in the system. Eliminating the feedback, we should find a way to derive DL-CSI using UL-CSI, which is the only available information about the environment at the base-station.

To better describe the problem, consider a block of time-frequency between a pair of transmitter-receiver antennas, as in Fig. \ref{fig:frame}. Assuming a grid over this block, the knowledge of the channel state information is equivalent to having information about the effects of the channel (on both amplitude and phase of the transmitted signal) over each cell of the grid (i.e., we should know a complex value for each grid cell). 

This block itself consists of two main portions of UL and DL: a) the first \(\Delta {f_{UL}}\) rows (subcarriers) of the frame are assigned to UL and b) the next \(\Delta {f_{DL}}\) rows (subcarriers) are assigned to DL. Columns of our frame represent different time slots (or how the channel effects change over time). By considering this structure, we can say that the problem at hand is having the UL-CSI information over the \(\Delta {f_{UL}}\) subcarriers and \(\Delta {t_{UL}}\) time slots (part 1) and wanting to predict DL-CSI in \(\Delta {f_{DL}}\) subcarriers and the next \(\Delta {t_{DL}}\) time slots (part 2). It is worth mentioning that to make the model realistic and causal, we only use the past UL-CSI information for DL-CSI prediction (and not the UL-CSI measured at the same time-slots of DL-CSI).

To solve DL prediction problem, most of the previous studies are based on first considering a mathematical channel model for the environment. For example, the \textit{multipath channel model} is defined as
\begin{equation} \label{eqn:multipath}
	h = \sum\limits_{n = 1}^N {{a_n}{e^{ - j2\pi \frac{{{d_n}}}{{{\lambda _0}}} + j{\phi _n}}}},
\end{equation}
where \(h\) is the channel response over particular frequency of \(\frac{1}{\lambda _0}\). In \eqref{eqn:multipath}, it is assumed that there are \(N\) distinct paths in the environment, where \(a_n\) is the path attenuation and \(\phi _n\) is a frequency-independent phase shift that captures reflection and attenuation of the signal along that path.


In a machine learning terminology, the common approach is that they first consider a parametric model for the environment and then use UL-CSI to estimate the parameters of the model. By obtaining the resulted complete model, the DL-CSI can be predicted.    

Incorrect assumptions about the parametric model and/or incorrect derivation of the parameters both lead to loss of some parts of UL-CSI information and consequently low accuracy of DL-CSI prediction. Furthermore, when parametric models are estimated, often some simplifying assumptions should be considered that may not be true in some cases or even violated. For example, as mentioned in section \ref{ssec:intro}, in Eq. \eqref{eqn:multipath}, \(a_n\) is assumed to be constant for UL and DL, but as some studies (e.g., \cite{han2018efficient}) suggest this assumption is not always correct.   

To avoid forcing incorrect latent domain and the problem of oversimplistic assumptions, in this work, we do not use any specific parametric model and instead use data-driven approaches to discover the underlying structure of data without any prior model assumptions. More details of the proposed scheme are presented in Section \ref{sec:proposedscheme}. 

\section{Proposed scheme} \label{sec:proposedscheme}

In this section, we first explain how CSI information can be considered as an image. Then, we present the two approaches, which we propose for  DL channel prediction.

\begin{figure}
	\centering
	\includegraphics[scale=0.4]{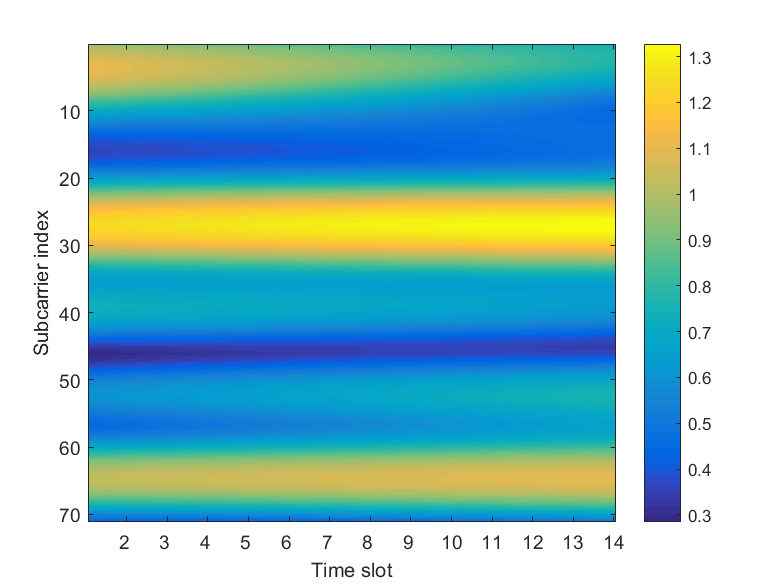}
	\centering
	\caption{CSI between a pair of antennas considered as an image}
	\label{fig:csiimage}
\end{figure}

\subsection{CSI as an Image}
Looking back at Fig. \ref{fig:frame}, CSI is a 2D complex matrix with size of \({N_s} \times {N_t}\), where \({N_s}\) is the number of subcarriers and \({N_t}\) is the number of time slots in the time-frequency block. 

Recently, many advanced techniques have been proposed for analyzing image data using neural networks. Image data are in fact 2D real matrices with one or more channels. For example, the color images are 2D images with 3 channels (for red, green, and blue components).  To use image-based techniques in this work we have considered the 2D CSI matrix as an image. For example in Fig. \ref{fig:csiimage}, the heat map of CSI absolute values are plotted for a sample FDD frame. 

If we use only absolute values, phase information will be lost. So to solve this problem, we consider the complex valued CSI matrix as a real valued matrix with two channels.
There are two choices to put complex values as two channels of an image: put absolute values in the first channel and phase values in the second channel or put real values in the first channel and imaginary values in the second channel. We selected the second approach to prevent the problem of phase wrapping that may happen for the phase information. 

In the rest of this paper  wherever we use the term "image", it refers to the CSI matrix that is considered as a \({N_s} \times {N_t} \times 2\) image with real values in the first channel and imaginary values in the second channel.  


\subsection{Up-Link to Down-Link Knowledge Transfer}
In this paper, to extract environment information from UL and then transfer the derived knowledge back to the DL domain, two approaches are introduced based on deep neural networks: direct approach and generative approach. In the following subsections we explain each of these two approaches in details.

\subsubsection{\textbf{Direct Approach}}

\begin{figure}[tbh!]
	\centering
	\includegraphics[scale=0.3]{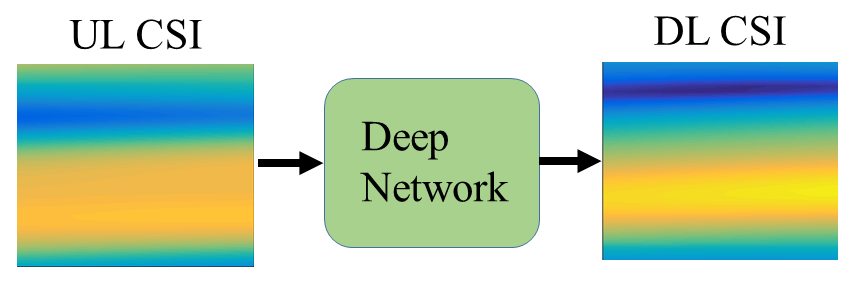}
	\centering
	\caption{Direct approach: considering Fig.~\ref{fig:frame}, the network gets "part 1" of the image (UL-CSI) as input, and tries to predict "part 2" of the image (DL-CSI) as output. Clearly, we do not use information of other parts of the image in this approach}
	\label{fig:direct}
\end{figure}

As mentioned before in UL to DL transfer, we have two steps: first encode the environment information from UL-CSI to a latent domain, second transfer and decode derived latent domain model into DL-CSI. In the direct approach we use a network to accomplish both of these two steps in a single process as one deep network (Fig. \ref{fig:direct}).

In this approach, we feed  UL-CSI in \(\Delta {f_{UL}}\) subcarriers and past \(\Delta {t_{UL}}\) time slots (part 1 in Fig. \ref{fig:frame}) as the input and the network tries to predict DL-CSI in \(\Delta {f_{DL}}\) subcarriers and next \(\Delta {t_{DL}}\) time slots (as shown in part 2 of Fig. \ref{fig:frame}) as the output.


As discussed in Section \ref{sec:background}, CNNs are among the most successful tools in analyzing image data, so in this work we design a specific convolutional neural network to implement direct approach and use the designed model to predict DL CSI. The details of the network structure will be discussed in Section \ref{sec:implementation}.

\subsubsection{\textbf{Generative Approach}}

We still have similar desired input and output: considering  \textit{part 1} of Fig. \ref{fig:frame} as the input and \textit{part 2} as the output. The difference is that we do not directly learn the UL to DL relation, instead, we consider the whole time-frequency block (a matrix of size  \({N_s} \times {N_t}\)) as the image that we want to learn, i.e., we want to learn the joint distribution between different pixels of the complete CSI image. Knowing the joint distribution of the whole  frame, we also know joint distribution of the UL section and DL section (Part 1 and part 2).  Using UL-DL joint distribution, we can  predict DL-CSI when we have the UL-CSI. It is clear that capturing such joint distribution is a very complicated task (considering the size of the CSI matrix).

As briefly described in Section \ref{sec:background}, researchers in the AI field, recently proposed the GAN structure as a very successful tool for estimating joint distribution of the input data, specially when we are dealing with images. After training a GAN with a set of images,  it can generate images that are very similar to the real images. \textit{One interesting application of GANs} is in image completion, i.e., given a corrupted image (like when one part of the image is missing) the GAN tries to find the missing part. Several schemes have been proposed for image completion. The core idea of them is that first GAN tries to generate an image that resembles (according to some metric) the corrupted image and then uses the generated image to predict the missing part.  

Motivated by the success of GANs and image completion schemes (and since we are able to consider the CSI matrix as an image) we should be able to use similar network structure to find the joint distribution of the CSI and then use that model to predict DL-CSI from UL-CSI. 

The steps of our proposed scheme can be summarized as:
\begin{itemize}
	\item \textbf{Training Phase:}
	First train a GAN with CSI images of the whole time-frequency block. After complete learning, the generator network is capable of getting a random vector $z$ as input and creating images, which are very similar to real CSI images, as shown in Fig. \ref{fig:GAN_gen}. 

\begin{figure}[tbh!]
	\centering
	\includegraphics[scale=0.4]{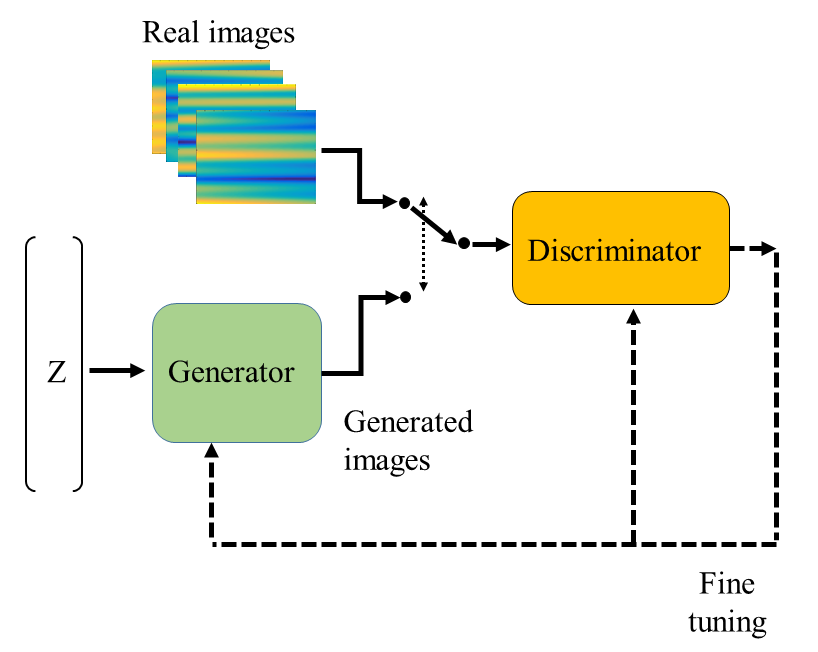}
	\centering
	\caption{Given a set of complete time-frequency CSI blocks, the generator learns how to create CSI images that are similar to the real CSI blocks}
	\label{fig:GAN_gen}
\end{figure}
   
	It is worth mentioning that from different types of GANs, we first selected the most common structure called Deep Convolutional Generative Adversarial Network (DCGAN) \cite{radford2015unsupervised}. Although DCGANs are able to produce similar images like our CSI images, during the completion phase, we did not get desired results and the MSE of prediction was relatively high. Additionally, given a UL-CSI frame, with different initializations of the input vector $z$ we got very different predictions of DL-CSI. 
		   
	To solve this problem, in this work, we have used BEGAN (described in Section \ref{ssec:GAN}). As discussed there BEGANs are designed based on the MSE error and have better convergence properties.
   
	\item 
	\textbf{Completion Phase:}
 
	In this step we want to predict the DL-CSI. {The idea} is that we consider \textit{the time-frequency block that only has the UL-CSI} as \textit{the corrupted image}, then we use different GAN-based image completion algorithms to complete the missing part (DL-CSI). As we treat the prediction task as completing a corrupted image, we name the prediction phase, as the completion phase, as shown in Fig. \ref{fig:completion}.
\begin{figure}[tbh!]
	\centering
	\includegraphics[scale=0.3]{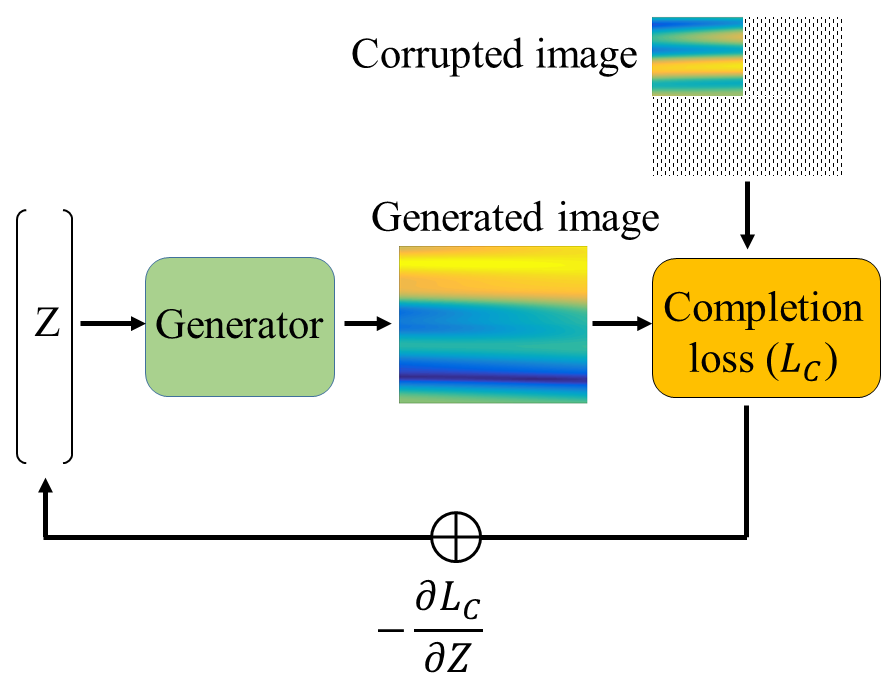}
	\centering
	\caption{Image completion procedure: given a corrupted image (only UL-CSI is known), the network generates appropriate image that can be used for DL-CSI prediction}
	\label{fig:completion}
\end{figure}

	More accurately, in the completion phase, the vector $z$ (input of the generator) is initialized with a random state. Then we update $z$ using the gradient descent method so that the generated image and the corrupted image become more similar (a loss function is reduced). After several iterations, the generated image is considered equal to a complete real image and the desired output (DL-CSI) will be derived. In this work we have used two different loss functions and tried image completion using both methods.

 \begin{figure*}
	\centering
	\begin{subfigure}{.5\textwidth}
		\centering
		\includegraphics[scale=0.28]{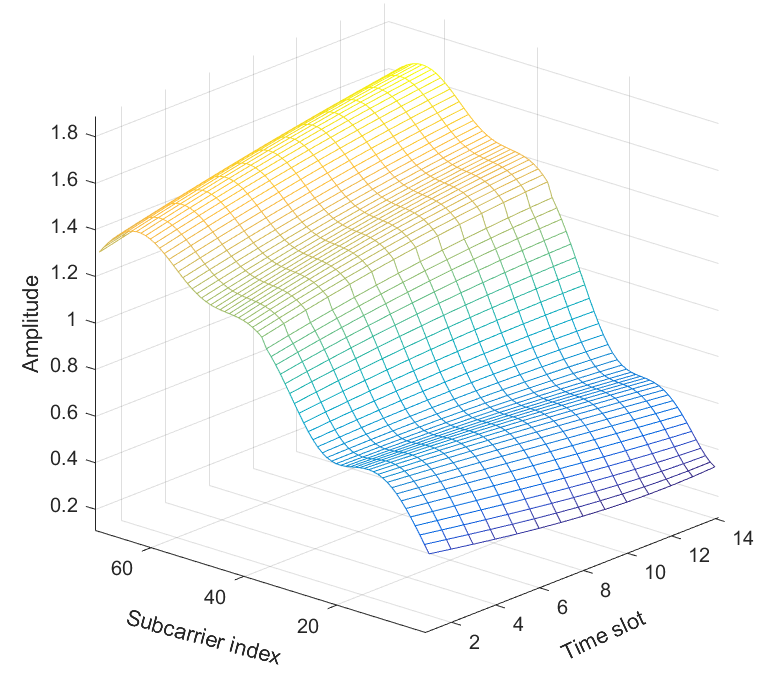}
		\caption{a sample EVA simulated frame}
		\label{fig:sub1}
	\end{subfigure}%
	\begin{subfigure}{.5\textwidth}
		\centering
		\includegraphics[scale=0.28]{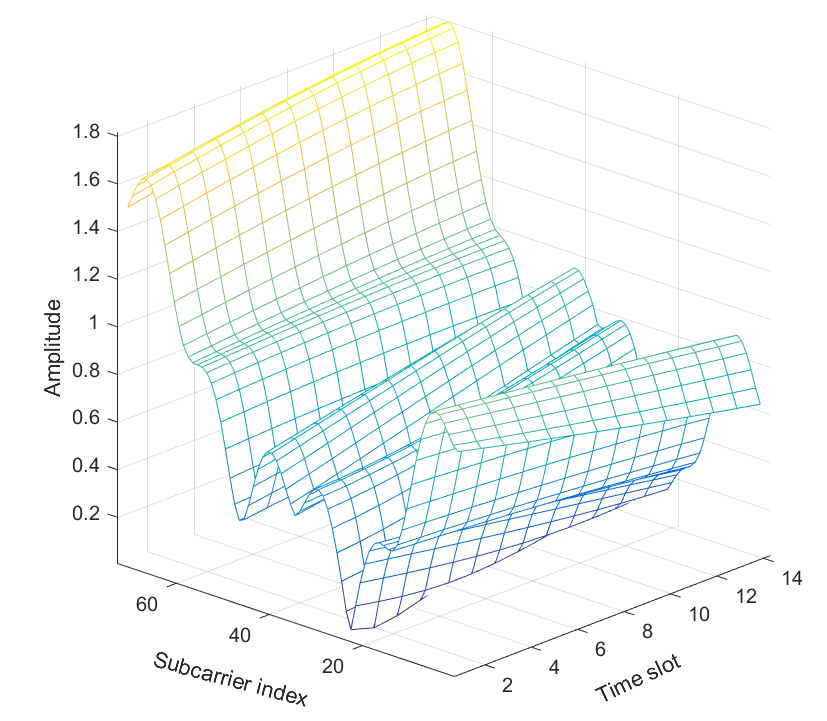}
		\caption{a sample ETU simulated frame}
		\label{fig:sub2}
	\end{subfigure}%
	\caption{Samples of time-frequency CSI block: EVA (\ref{fig:sub1}) and ETU (\ref{fig:sub2}) }
	\label{fig:datasetsampel}
\end{figure*}

\paragraph{\textbf{Contextual Loss}}
Contextual loss is defined as the distance between a known part of the image and its corresponding part in the generated image. If we define the mask as

 
	\begin{equation}
		\mbox{mask}\left[ {i,j} \right]=
		\begin{cases}
		1,\;\; & 0 \le i \le \Delta {f_{UL}},0 \le j \le \Delta {t_{UL}}\\
		0,\;\; & \mbox{otherwise}
		\end{cases}.
	\end{equation}
	 
	Then the contextual loss will be
	\begin{equation}
		\mbox{contextual loss} = \left\| {\mbox{mask} \odot x - \mbox{mask} \odot G\left( z \right)} \right\|,
	\end{equation}
	where \textit{x} is the image that we want to complete and \(G\left( z \right)\) is the generated image.
	  	
	\paragraph{\textbf{Contextual Loss + Perceptual Loss}}
	
	Such loss function for image completion was first used in \cite{yeh2017semantic}. If we use only contextual loss, the final completed image may seem artificial (having different structure compared to real data) so one uses discriminator loss of generated image as a new term in the total loss and calls that perceptual loss, since it gives a sense of being real. Hence, in BEGAN
	\begin{equation}
	\mbox{perceptual loss} = D\left( {G\left( z \right)} \right),
	\end{equation}
	and,
	\begin{equation}
		\mbox{total loss} = \mbox{contextual loss} + \lambda\times \mbox{perceptual loss},
	\end{equation}
	where \(\lambda\) is a hyper parameter to control how much emphasis is put on the perceptual loss with respect to contextual loss while performing the gradient descent. Its default value in BEGAN is 0.01.

		  
\end{itemize}

\section{Implementation and Simulation Results} \label{sec:implementation}

In the following, we discuss the details of the implementation and the simulation results. The source code of the implementation can be found at {https://github.com/safarisadegh/UL2DL}


\begin{figure*}
	\centering
	\includegraphics[scale=0.5]{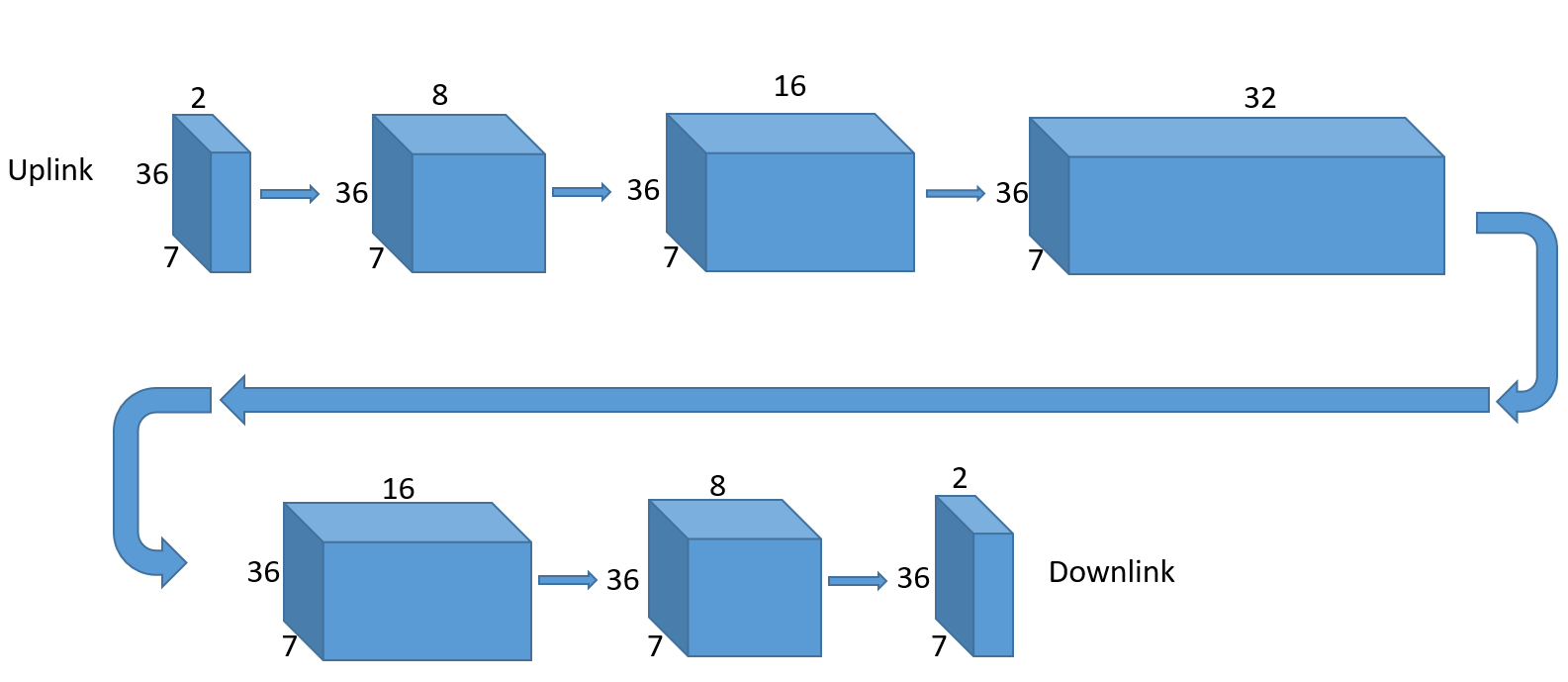}
	\centering
	\caption{CNN structure used in the direct approach}
	\label{fig:cnn}
\end{figure*}

\subsection{Dataset Generation}

To evaluate the performance of our proposed schemes, we have used Vienna LTE-A Downlink link level simulator \cite{mehlfuhrer2011vienna} to simulate multipath fading channels. Two 3GPP fading models were simulated for a single-input single-output (SISO) channel: Extended Vehicular A (EVA) and Extended Typical Urban (ETU) and we used speed of $50$ km/h to take into account the Doppler effect. Our simulated time-frequency frames had size of $72$\(\times\)$14$ ($72$ subcarriers in $14$ time slots equivalent to $6$ resource blocks in a $1$ ms subframe). These numbers are selected to have similarity to the 3GPP LTE FDD frame structure. As for simulations, we select the first $36$ subcarriers over the first $7$ time slots as the UL channel, and the second $36$ subcarriers over the second $7$ time slots as the DL channel. 

\begin{figure*}
	\centering
	\begin{subfigure}{.5\textwidth}
		\centering
		\includegraphics[scale=0.5]{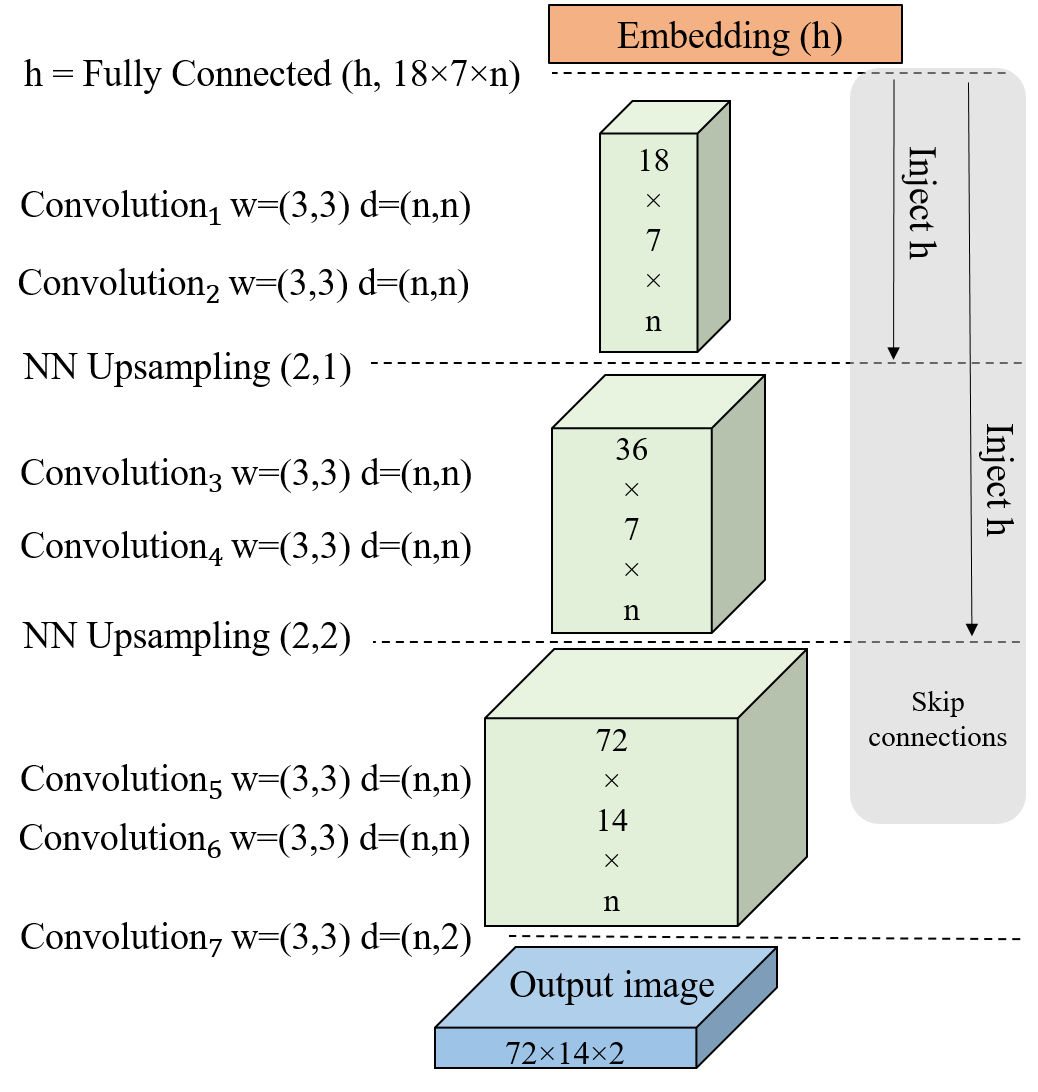}
		\caption{decoder (generator)}
		\label{fig:decoder}
	\end{subfigure}%
	\begin{subfigure}{.5\textwidth}
		\centering
		\includegraphics[scale=0.5]{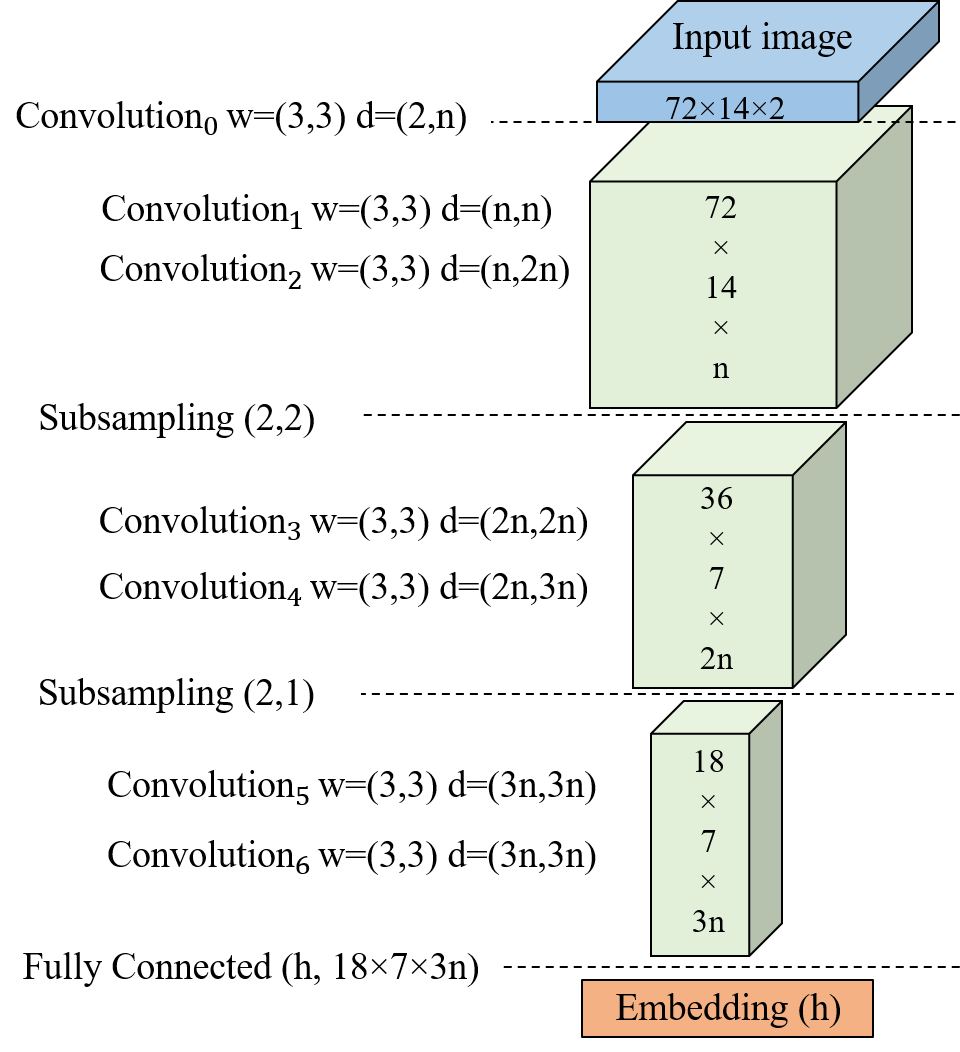}
		\caption{encoder}
		\label{fig:encoder}
	\end{subfigure}%
	\caption{BEGAN structure used in the generative approach (w: kernel size, d: input and output channel dimensions of the layer, n: a hyperparameter for hidden layers, which we set to n = $64$)}
	\label{fig:beganstructure}
\end{figure*}



The number of simulated frames that were created independently was $40$K ($35$K for training, $2$k for validation and $3$K for test). Samples of simulated frames are shown in Fig. \ref{fig:datasetsampel}. For convenience, we only show absolute values of frames (We note that for training/testing of the networks we always feed CSI as real and imaginary parts, and just for the presentation we show the CSI absolute value; the good match between the predicted and actual absolute values  means good prediction on both real and imaginary parts).

\subsection{Network Structure}


\begin{figure*}
	\centering
	\includegraphics[scale=.8]{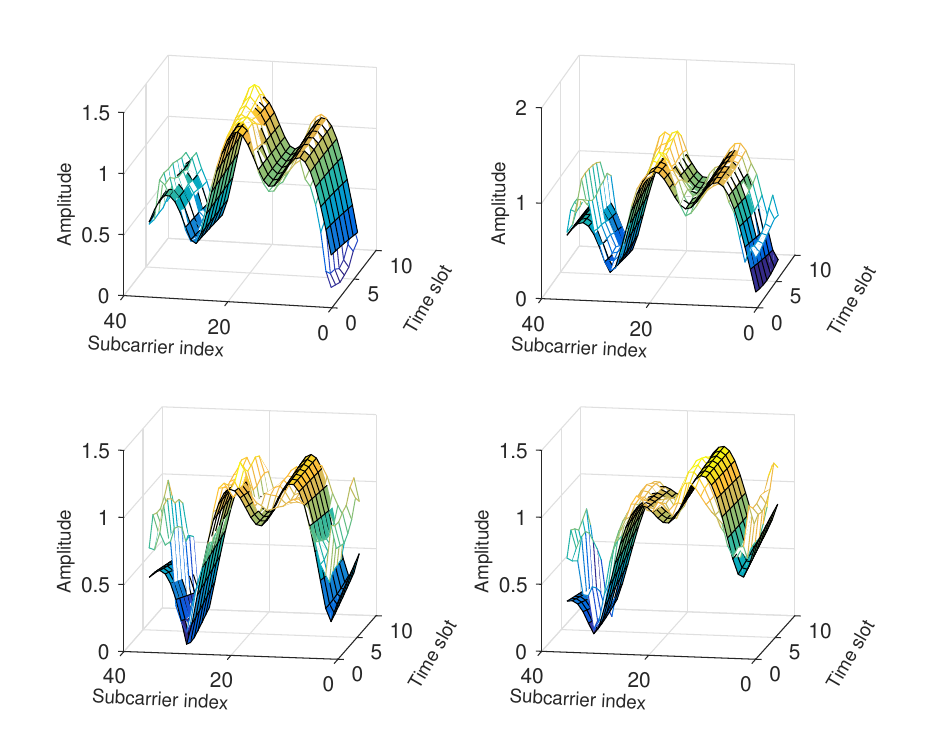}
	\caption{Samples of DL-CSI prediction using the direct approach on ETU dataset with ground truth depicted as a surface (solid face color) and predicted DL-CSI depicted as a meshgrid plot.}
	\label{fig:cnntest_etu}
\end{figure*}

\begin{figure*}
	\centering
	\includegraphics[scale=.8]{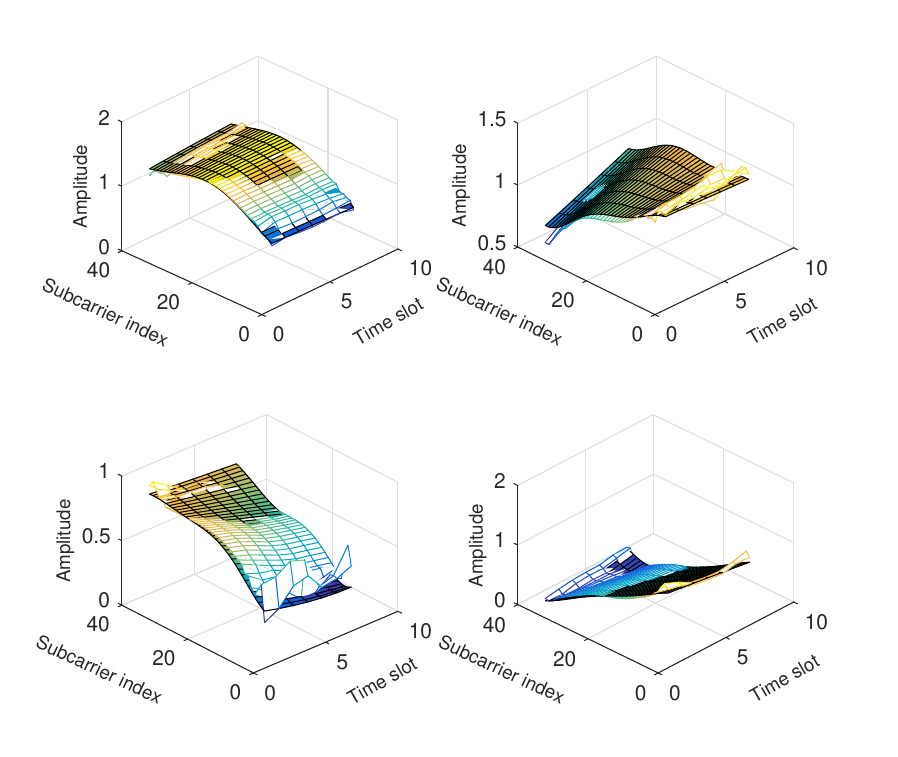}
	\caption{Samples of DL-CSI prediction using the direct approach on EVA dataset with ground truth  depicted as a surface (solid face color) and predicted DL-CSI depicted as a meshgrid plot.}
	\label{fig:cnntest_eva}
\end{figure*}

In this section we explain the network structure that we have used for DL-CSI prediction for each of the direct and generative approaches.

\begin{figure*}
	\centering
	\begin{subfigure}{.5\textwidth}
		\centering
		\includegraphics[scale=0.6]{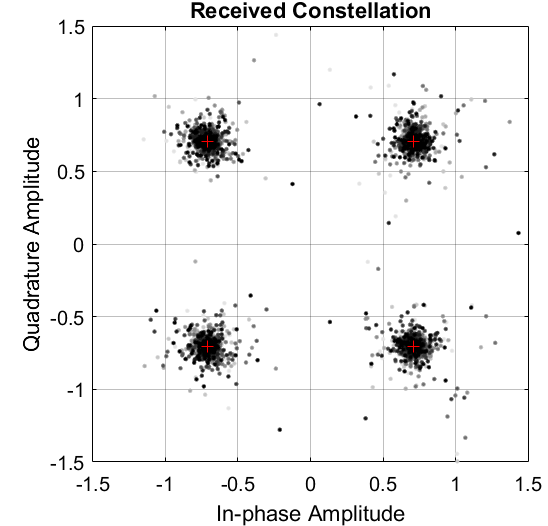}
		\caption{EVA model}
		\label{fig:constelcnnsub1}
	\end{subfigure}%
	\begin{subfigure}{.5\textwidth}
		\centering
		\includegraphics[scale=0.6]{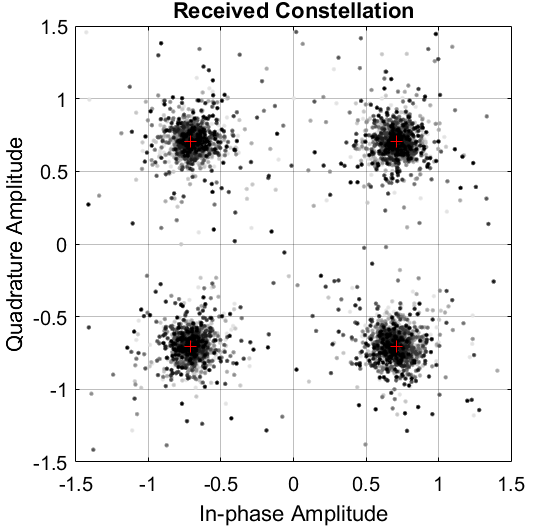}
		\caption{ETU model}
		\label{fig:constelcnnsub2}
	\end{subfigure}%
	\caption{{Received constellations for $5$K pre-compensated QPSK symbols using DL-CSI prediction (direct approach)}}
	\label{fig:constelcnn}
\end{figure*}

\subsubsection{Direct Approach}
As mentioned before, CNN is used to implement the direct approach. The designed network only contains convolutional layers (there is no pooling or fully connected layers) that results in lower training and testing complexity. We aimed to design our network as simple as possible thus it has only $5$ hidden layers. It has a total of about $12$K learnable parameters that is very small compared to a few million parameters in typical deep networks. \(\tanh\) activation function is used in all layers of the network except the output layer (\(lrelu\) activation function is also tested but \(\tanh\) results in better predictions). The deployed network structure is shown in Fig. \ref{fig:cnn}.

We used Xavier method \cite{glorot2010understanding} to initialize network parameters. For optimization, we used Adam optimizer \cite{kingma2014adam}. Except for the first two layers, which have symmetric padding, we used zero padding for other layers.

\subsubsection{Generative Approach}
To adapt the network structure to our image size (\(72 \times 14 \times 2\)), we used main BEGAN \cite{berthelot2017began} structure with some modifications. The network structure is shown in Fig. \ref{fig:beganstructure}. CSI values also normalized to their maximum value and during training procedure a zero mean normal noise with decaying variance was added to input values to improve network regularization. Other settings are similar to \cite{berthelot2017began}. 

During the training phase we also faced the mode collapse problem. In mode collapse, the generator generates one or limited sets of images. It is a common problem in training GANs and as mentioned in \cite{berthelot2017began} can be seen in BEGANs. Berthelot et al. \cite{berthelot2017began} suggest that decreasing the start learning rate can help the network recover from the mode collapse issue but in our case it does not help us to recover from the collapse mode. The solution we have used in this paper is to share weights between the generator and the decoder part of the discriminator,
(i.e., not only the generator and the decoder part of the discriminator/  generator have the same structures, but also we have used the same weights for both of them, in contrast to \cite{berthelot2017began}, where they could have different weights).

\subsection{Simulation Results}

\subsubsection{\textbf{Direct Approach}}
After we trained our CNN on EVA and ETU datasets, we used the trained CNN to predict DL-CSI.~\textcolor{black}{Some samples of the predicted DL-CSI and their corresponding ground truth (matrices of shapes $36\times 7$) are shown in Fig. \ref{fig:cnntest_etu} and Fig. \ref{fig:cnntest_eva}.  The actual and predicted DL-CSI are depicted as surface (solid face colors) and meshgrid plots, respectively.  As can be seen, the network performance on EVA dataset is better than ETU dataset (for ETU dataset most errors occurred on edge subcarriers).} 

Despite the good prediction quality, as can be seen in Fig. \ref{fig:cnntest_etu}, sometimes we have relatively large errors at the edge subcarriers. This is due to the structure of convolutional layers (mainly due to the lack of fully connected layers). One way to correct the edges is to train the network on  images larger than $36\times 7$ and then look at the middle $36\times 7$ block.

\textbf{Discussion on the CNN Structure}:
Kernel size is one of the most important hyperparameters in convolutional networks. Several kernel sizes have been examined in this work where $3\times 3$ leads to the best results. Although we cannot mathematically prove, we might be able to say that when we use smaller filter sizes, the receptive fields of the filter become smaller and thus it would be easier to detect local features in the image (data). As for the channel image, since the channel values are more locally correlated (and they are more independent when they are farther from each other), smaller filter sizes should be a better choice. On the other hand, $1\times 1$ kernel size is not adopted as it could not look into the relationship between the neighboring pixel information.

Furthermore, in typical CNNs, in last layers of the network, there are one or two fully connected layers but in this work to reduce the complexity of the network we did not use any fully connected layers. Addition of these layers would considerably increase the number of trainable parameters, which makes the training procedure more complicated. For example, if we add a fully connected layer as the last layer of the network, there will be about 1 million additional parameters while currently, our CNN network has only $12$K parameters.

We also did not use pooling or large strides in our convolution layers. This has two main reasons:
\begin{enumerate}
\item DL-CSI (output) is of the same dimensions as the UL-CSI (input), so the stride is set at 1 to keep the dimensions, and we use the same padding for the missing boundary pixels.
\item  In typical CNN's, by use of pooling or stride convolution the network can be oblivious to rotations and shifts in the input image. For CSI images, however, rotation and shift are sources of amplitude and phase distortion and obviously should not be discarded. 
\end{enumerate}



To see the performance of the predicted DL-CSI, let us consider a pair of a transmitter (base-station) and a receiver (user). Also assume that the user and the signaling need to be simple so the user is not able to estimate the channel and feed it back to the server. In such settings if the base-station wants to send data to the user, it needs to pre-compensate the transmitted data.

To perform the pre-compensation, the base-station needs to know the DL-CSI but since there is no feedback, it should predict that. Therefore, the procedure for the transmitter is to first measure the UL-CSI and then use the proposed scheme to predict the DL channel state in the next time slot. Having a prediction of the DL channel it can pre-compensate the signal.

\begin{figure*}
	\centering
	\begin{subfigure}{1\textwidth}
		\centering
		\includegraphics[scale=0.37]{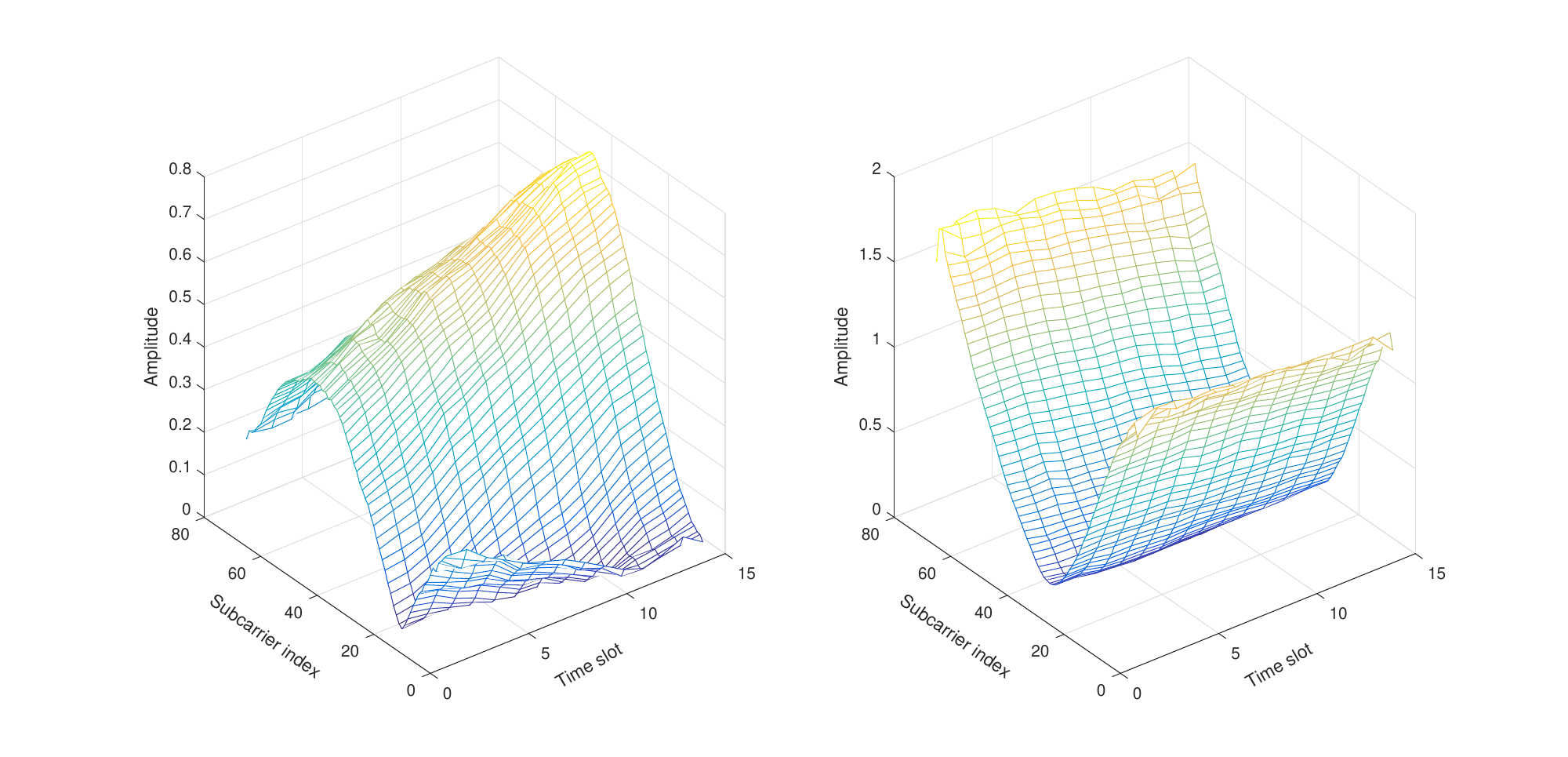}
		\caption{EVA model}
		\label{fig:gansub2}
	\end{subfigure}
	\begin{subfigure}{1\textwidth}
		\centering
		\includegraphics[scale=0.37]{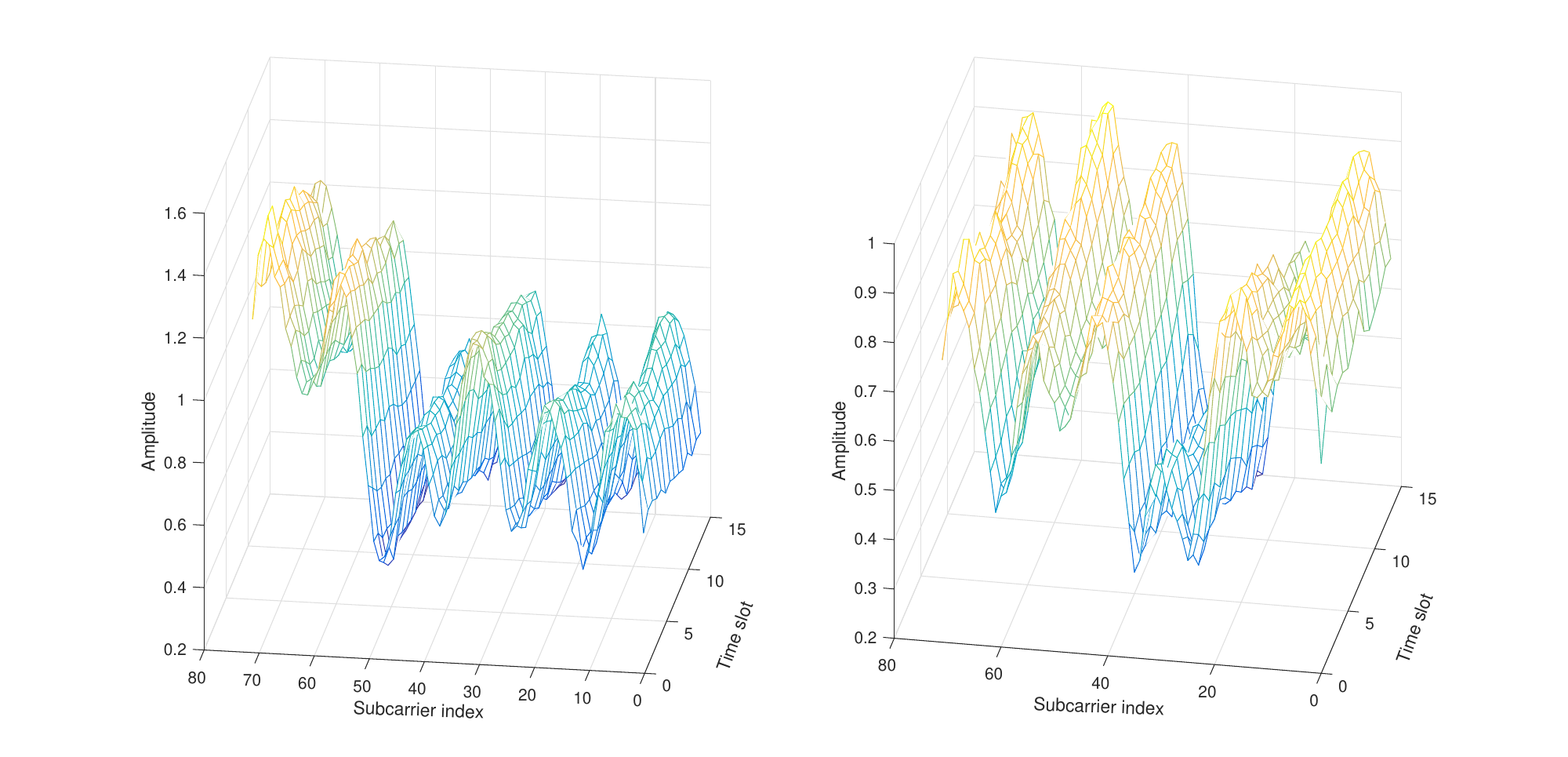}
		\caption{ETU model}
		\label{fig:gansub3}
	\end{subfigure}
	\caption{Samples of the generated images}
	\label{fig:gantest}
\end{figure*}

To examine such settings, we have simulated many channel realizations (both UL and DL channels). For each case, we assume that the base-station only knows the UL channel and it uses that information to predict the DL channel. Assuming that the base-station wants to send QPSK modulated signals, before transmission it divides the signal by what it has \textit{predicted for the DL channel}. The pre-compensated signal is then transmitted through the downlink (and thus will be multiplied by the \textit{actual realization of the DL channel}). Therefore, if we have a good prediction of downlink they will cancel out each other.   Constellations of the received symbols are shown in Fig. \ref{fig:constelcnn} for EVA and ETU channel models. As can be seen, the received constellations are well concentered around the QPSK points verifying high prediction accuracy.

\begin{figure*}
	\centering
	\begin{subfigure}{1\textwidth}
		\centering
		\includegraphics[scale=0.37]{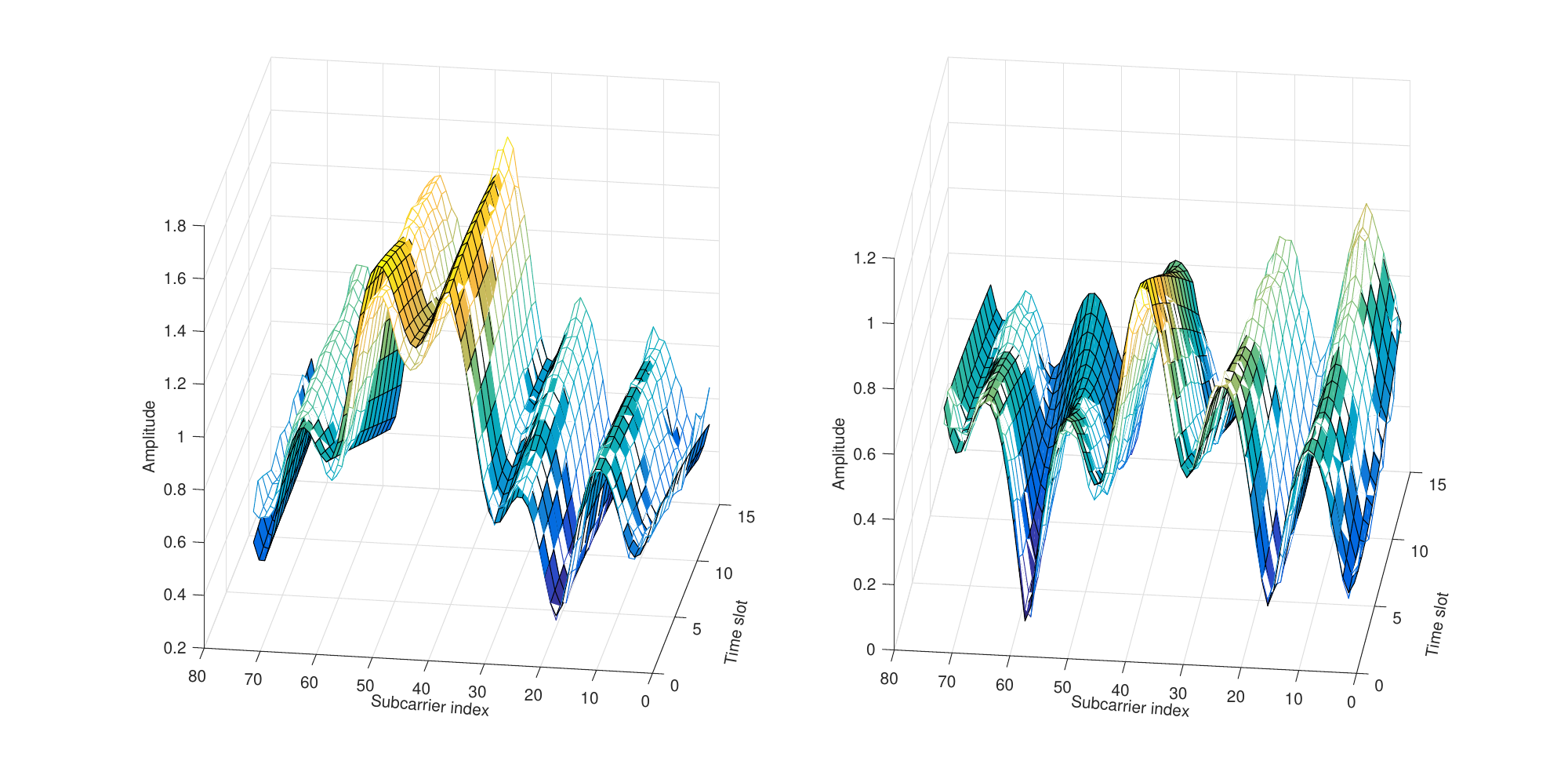}
		\caption{ETU model}
		\label{fig:compsub1}
	\end{subfigure}
	\begin{subfigure}{1\textwidth}
		\centering
		\includegraphics[scale=0.37]{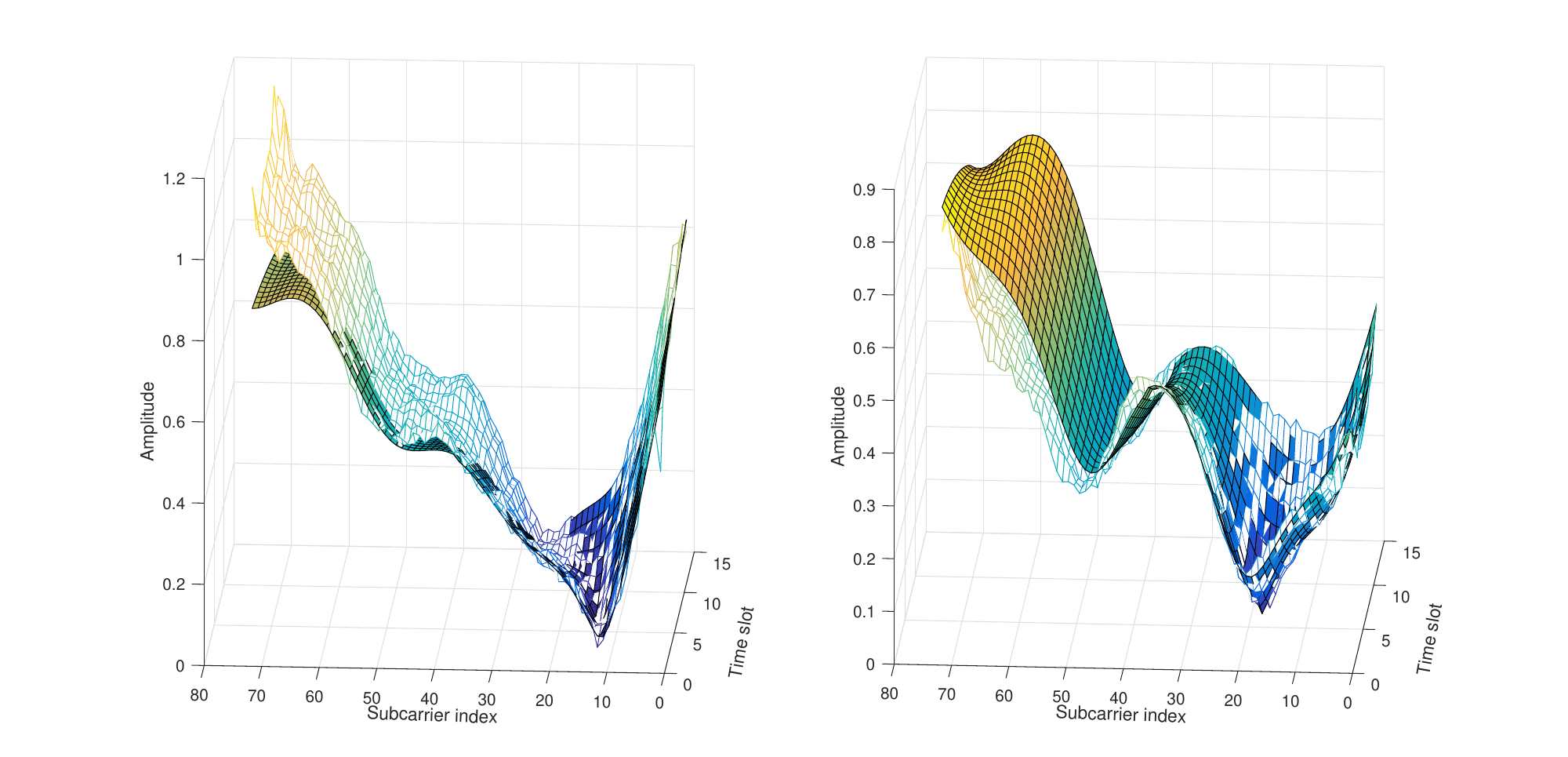}
		\caption{EVA model}
		\label{fig:compsub3}
	\end{subfigure}
	\caption{ Samples of image completion using the generative approach with the ground truth depicted as a surface of solid face colors and generated CSI blocks depicted as a meshgrid.}
	\label{fig:comptest}
\end{figure*}

\begin{figure}
	\centering
	\includegraphics[scale=0.37]{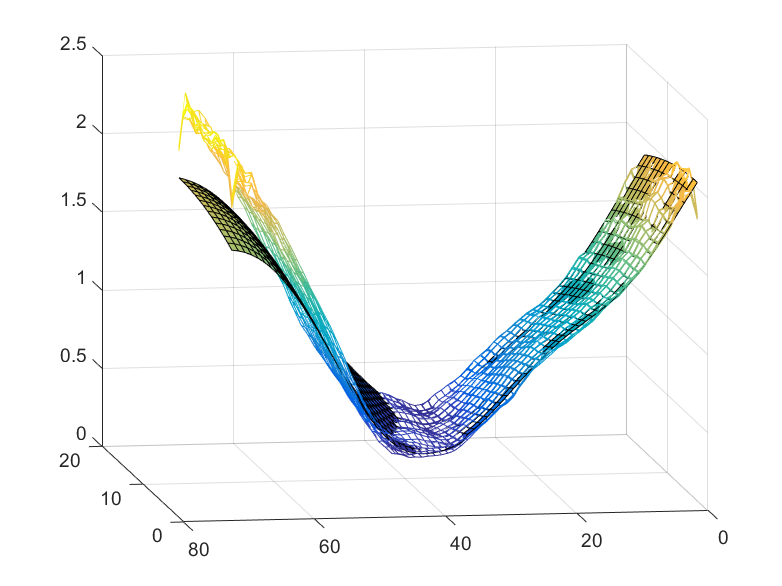}
	\centering
	\caption{Completion results with different initializations of the $z$ vector: completion is performed for 5 different initialized $z$ vectors. As shown, the results are almost the same.}
	\label{fig:samecomp}
\end{figure}

\subsubsection{\textbf{Generative Approach}}
We repeat the above studies to see the performance of the second proposed scheme. 

First, we trained BEGAN on EVA and ETU datasets to generate images like complete frequency-time CSI block. Some BEGAN generated images are shown in Fig. \ref{fig:gantest}. Note that these are images of size $72\times 14$ (the whole time-frequency block) not just the DL-CSI.

\begin{figure*}
	\centering
	\begin{subfigure}{.5\textwidth}
		\centering
		\includegraphics[scale=0.6]{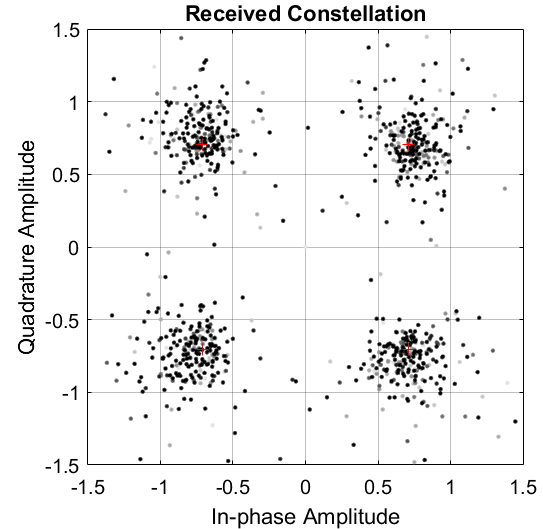}
		\caption{ETU model}
		\label{fig:constelgansub1}
	\end{subfigure}%
	\begin{subfigure}{.5\textwidth}
		\centering
		\includegraphics[scale=0.6]{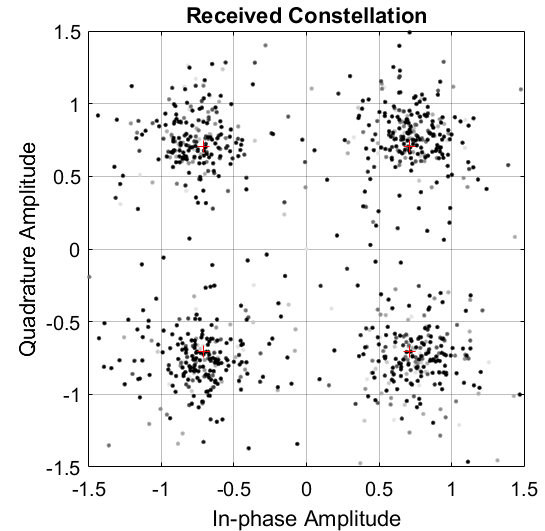}
		\caption{EVA model}
		\label{fig:constelgansub2}
	\end{subfigure}%
	\caption{Received constellations for 1K pre-compensated QPSK symbols using DL-CSI prediction (generative approach)}
	\label{fig:constelgan}
\end{figure*}

During the completion phase, the trained BEGAN is used to restore the missing parts in the CSI image, where UL-CSI is known and other parts are missing. The DL-CSI part of the resulting generated CSI image is then considered as the DL-CSI. Some completion examples using contextual loss are shown in Fig. \ref{fig:comptest} (different losses do not have notable visual differences, so we do not include them here; the numerical result is reported at the end though). In Fig. \ref{fig:comptest}, the actual time-frequency block of size \(72 \times 14\) is depicted as a surface (solid face colors) and the generated image is shown as a meshgrid. The DL-CSI relates to subcarriers $36$ to $72$ and time slots $7$ to $14$.

We have also tested the consistency of the DL-CSI image prediction, meaning that we fixed the UL-CSI part and then executed the image completion algorithm with different initializations of the $z$ vector to produce the complete image.
We note that one of the main problems we faced in DCGAN structure was the large difference between completed images (for a fixed corrupted image) for different initializations of $z$ vector. As seen in Fig.~\ref{fig:samecomp} this problem is solved using BEGAN. The ground truth is shown as a surface and the generated images are depicted as meshgrids (there are five generated images but as they are very close they are not easily differentiable).


To see the performance of this approach, we followed the same procedure as the direct approach and simulated the constellation map of
the received signal when we perform signal pre-compensation using the predicted DL-CSI. Resulted constellations are shown in Fig. \ref{fig:constelgan} for EVA and ETU channel models.

\subsubsection{\textbf{Comparing the Results}}

As in the last section, we present the comparative results between the performance of different proposed DL-CSI estimation methods. As for the comparison metric, we have used normalized mean squared error (NMSE), which is defined as:

\begin{equation}
NMSE = \frac{{{{\left\| {H - \hat H} \right\|}^2}}}{{{{\left\| H \right\|}^2}}},
\end{equation}
where \(H\) is the ground truth DL-CSI, and \(\hat H\) is the predicted value.


\xdef\tempwidth{\the\linewidth}
	
\begin{table} [!h]
	\centering
	\begin{tabular}{ccccc} 
		\hline
		Dataset & \multicolumn{2}{r}{Error}                      \\\hline
		& CNN    & \multicolumn{3}{l}{\kern 3pc  BEGAN}             \\ \hline
		&        & Contextual &  & Contextual+Perceptual \\ \hline
		EVA     & 0.0102 & 0.0638     &  & 0.0632                \\ \hline
		ETU     & 0.0376 & 0.0297     &  & 0.0308                \\[1ex] \hline
	\end{tabular}
	\caption{NMSE of channel predictions for different methods on EVA and ETU datasets}
\label{table:res}
\end{table}



In general, we anticipate the CNN performance to be superior on both datasets, since it has been trained directly to reduce DL-CSI MSE prediction, whereas, in the generative approach (BEGAN) the network is trained to capture CSI distribution (not the MSE of DL-CSI prediction). Based on the results of Table \ref{table:res}, the performance of CNN (as expected) is better on EVA dataset; however, BEGAN's performance is better on ETU dataset, which is a more complex environment. This observation may lead us to the conclusion that advanced techniques (like GAN networks) are more suitable 
for channels with higher complexity.
In BEGAN, both contextual and contextual plus perceptual losses have almost similar performances. On the EVA dataset contextual plus perceptual loss is slightly better while on the ETU dataset contextual is slightly better. This may be the indication that BEGAN learns the distribution correctly, and the ambiguity of the image completion phase is not large, and with both losses, we are able to determine the correct image.


It is worth mentioning that previous studies (e.g., AOA-based methods, covariance-based methods, and path extraction-based methods) that aim to predict DL-channel based on UL-channel focus on finding the MIMO channel matrix. In MIMO channel matrix each pair of Tx-Rx antenna is represented by only one value. In other words, those studies do not consider the time-frequency response that we have investigated in this paper. Therefore, it is not possible to compare the results of the proposed scheme with what has been suggested previously. 
There also exist some pilot-based schemes for DL-channel time-frequency estimation, e.g., minimum mean square error (MMSE) and least squares (LS), but since in our method there is no downlink pilot they are not suitable for comparing as a baseline.

\textcolor{black}{	
\subsection{Further Discussion}
\subsubsection{\textbf{MIMO Extension}}
To show how the proposed scheme can be used in MIMO settings, without loss of generality, we evaluated our methods on $2 \times 2 $ MIMO EVA and ETU channel models.}

\begin{figure*}
	\centering
	\includegraphics[scale=0.8]{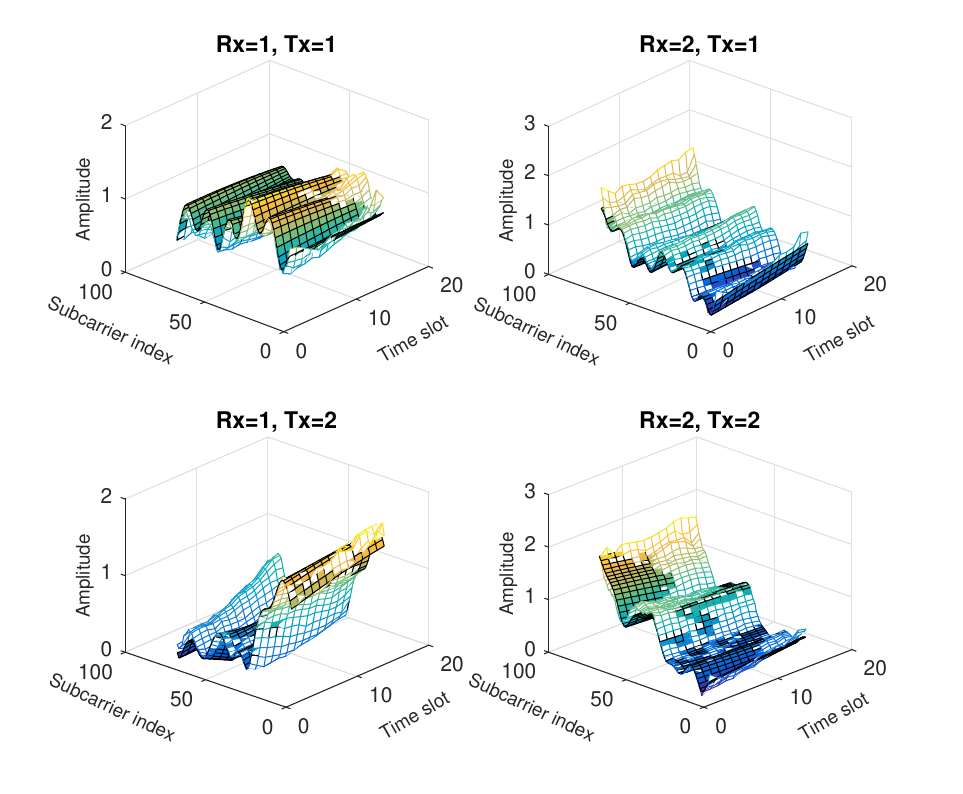}
	\centering
	\caption{\textcolor{black}{Samples of DL-CSI predictions using the generative approach (mesh grid) vs. the ground truth (solid face color) for a $2 \times 2$ ETU MIMO channel}}
	\label{fig:etugan_mimo}
\end{figure*}

\textcolor{black}{In previous sections we showed how to train a network to predict the DL channel from the UL channel for a single Tx-Rx link. Now we consider the $N_R \times N_T$ MIMO channel (with $N_R$ receiver antennas and $N_T$ transmitter antennas) as $N_R \times N_T$ Tx-Rx links, each link representing a pair of transmit receive antennas. We then feed our network with the uplink part of each Tx-Rx antenna pair, and the network outputs the prediction of the downlink part. The NMSE between the actual and predicted DL channels is considered as the performance metric.}

\begin{figure*}[tbh!]
	\centering
	\includegraphics[scale=0.8]{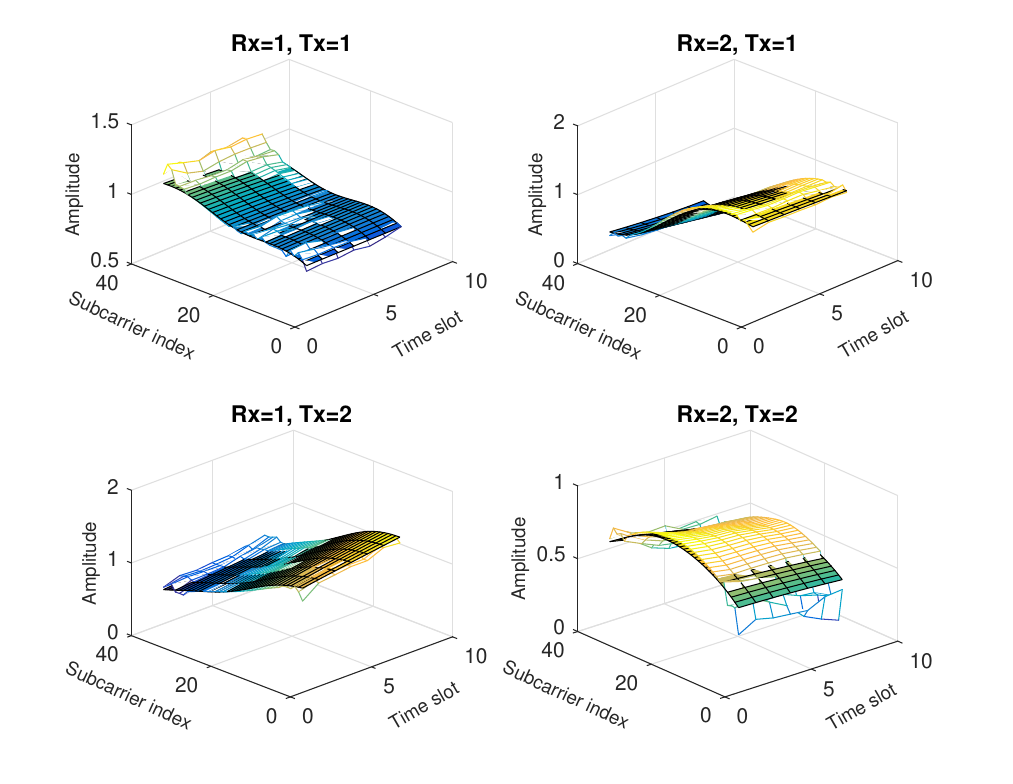}
	\centering
	\caption{\textcolor{black}{Samples of DL-CSI predictions using the direct approach (meshgrid) vs. the ground truth (solid face color) for a $2 \times 2$ EVA MIMO channel}}
	\label{fig:evacnn_mimo}
\end{figure*}

\textcolor{black}{
For example, Table \ref{table:mimo} compares the NMSE of predictions for BEGAN on ETU channel model and CNN on EVA channel model. As can be seen, training of one single network on a single antenna pair  (as we proposed) is enough to fully characterize a MIMO network on all dimensions of transmit antennas, 2) receive antennas, 3) time slots, and 4) frequency blocks. \textcolor{black}{Fig. \ref{fig:etugan_mimo} and Fig. \ref{fig:evacnn_mimo} each present $2\times 2$ DL-CSI prediction samples for ETU and EVA channel models, respectively.} 
\begin{table}[!h]
	\centering
	\begin{tabular}{ccc}
		\hline
		(Tx, Rx) & \multicolumn{2}{c}{Error} \\ \hline
		& CNN-EVA    & BEGAN-ETU    \\ \hline
		(1, 1)   & 0.0108     & 0.0359       \\ \hline
		(1, 2)   & 0.0102     & 0.0325       \\ \hline
		(2, 1)   & 0.0101     & 0.0348       \\ \hline
		(2, 2)   & 0.0101     & 0.0334       \\ \hline
	\end{tabular}
	\caption{\textcolor{black}{NMSE of prediction results for CNN and BEGAN methods for a $2 \times 2 $ MIMO channel}}
\label{table:mimo}
\end{table}}

\begin{figure*}
	\centering
	\includegraphics[scale=0.8]{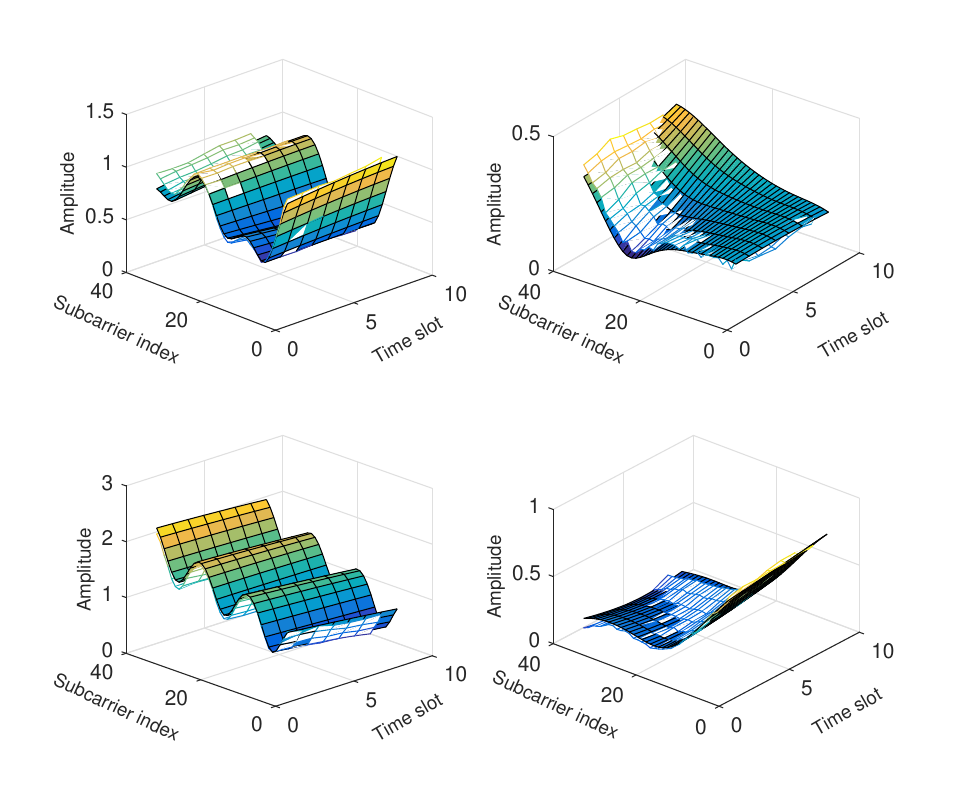}
	\centering
	\caption{\textcolor{black}{Performance of the network trained on both ETU and EVA channel models simultaneously. The ground truth is depicted as solid face color and the prediction as a meshgrid. }}
	\label{fig:comb_evaetu}
\end{figure*}

\textcolor{black}{
\subsubsection{\textbf{Various Channel Models}}
In the previous sections we developed separate networks for different channel models, e.g., ETU and EVA. So, one question that might arise is that if a single network can be used for different channel models. For this purpose, we have concatenated EVA and ETU simulated frames and trained one network that can handle both models simultaneously. Fortunately, the network successfully learned to handle both models simultaneously and the average prediction error of the network was between errors of the previous two separate networks. This shows that if we have enough samples and a network with enough parameters the network can handle different models concurrently. \textcolor{black}{A few samples of the actual and reconstructed DL portions of the channel are shown in Fig. \ref{fig:comb_evaetu}}}.  

\textcolor{black}{
\subsubsection{\textbf{Various User Speeds}}
 During training, EVA and ETU models were trained with samples in which the user speed was $50~\mathrm{km/h}$. However, to investigate the effects of varying speeds of users, we tested our networks (trained using samples of users with the speed of $50~\mathrm{km/h}$) to observe the prediction accuracy of the DL channel with different speeds. As can be seen from Fig. \ref{fig:speeds}, trained networks still accurately predict DL channels even with different user speeds during test time.
\begin{figure}
	\centering
	\includegraphics[scale=0.6]{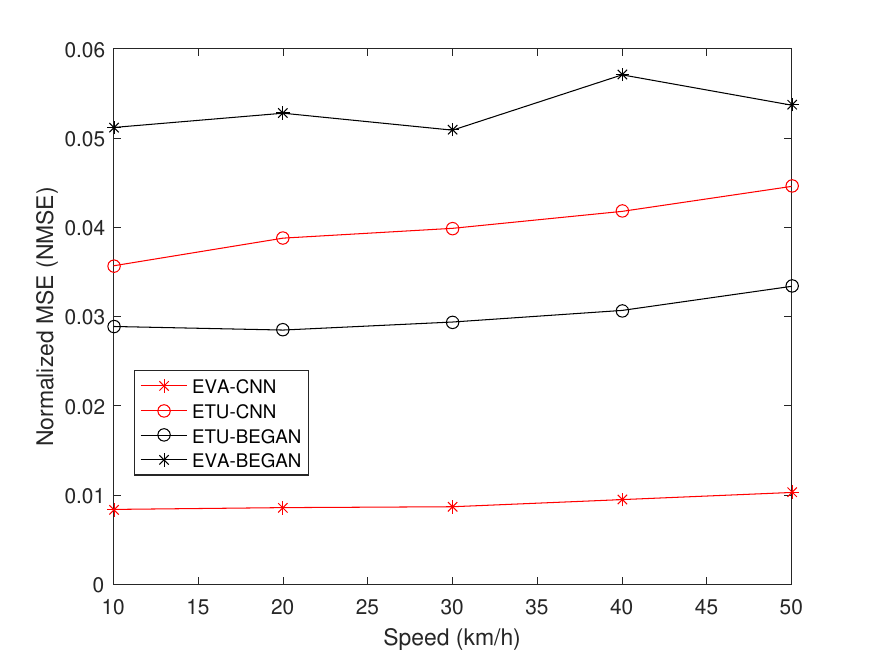}
	\centering
	\caption{\textcolor{black}{
		Channel prediction performance of the proposed models trained with the speed of $50~\mathrm{km/h}$ and tested at various other speeds}}
	\label{fig:speeds}
\end{figure}
}

\section{Conclusion} \label{sec:conclusion}
In this paper, we have proposed two data-driven approaches to predict DL-CSI from UL-CSI in FDD systems: direct approach and generative approach. Both of the proposed approaches try to use UL-CSI to determine a latent model that represents the environment propagation properties. The latent model is then used to predict DL-CSI. To determine the latent model, we have used convolutional neural network and generative adversarial network architectures for the direct and generative approaches, respectively. Our simulation results on EVA and ETU channel models show that even with simple neural networks we can predict DL-CSI (based on the observation of the UL-CSI), however, for environments with complex multipath structures we need more complex networks.


\section{Acknowledgment} \label{sec:Ack}
\textcolor{black}{We would like to thank anonymous reviewers for their help in improving the quality of this manuscript.}

\bibliographystyle{ieeetr}
\bibliography{ref}

\end{document}